\documentclass[3p,times,procedia]{elsarticle}
\usepackage{nupha_ecrc}

\usepackage{hyperref}
\usepackage{lineno}
\usepackage{graphicx,subfigure}
\usepackage{color}
\definecolor{dgreen}{cmyk}{1.,0.,1.,0.2}        
\definecolor{orange}{cmyk}{0.,0.353,1.,0.}    

\volume{00}

\firstpage{1}

\journalname{Nuclear Physics A}

\runauth{You Zhou (on behalf of the ALICE Collaboration)}

\jid{nupha}

\jnltitlelogo{Nuclear Physics A}

\usepackage{amssymb}

\usepackage{lineno}

\usepackage[figuresright]{rotating}

\begin{document}

\begin{frontmatter}




\title{Probing non-linearity of higher order anisotropic flow in Pb--Pb collisions }
\author{You Zhou (on behalf of the ALICE Collaboration)}
\ead{you.zhou@cern.ch}
\address{Niels Bohr Institute, University of Copenhagen, Blegdamsvej 17, 2100 Copenhag1en, Denmark}




\begin{abstract}
The second and the third order anisotropic flow, $V_{2}$ and $V_3$, are determined by the corresponding initial spatial anisotropy coefficients, $\varepsilon_{2}$ and $\varepsilon_{3}$, in the initial density distribution. On the contrary, the higher order anisotropic flow $V_n$ ($n > 3$), in addition to their dependence on the same order initial anisotropy coefficient $\varepsilon_{n}$, have a significant contribution from lower order initial anisotropy coefficients, which leads to mode-coupling effects. In this contribution, we present the investigations on linear and non-linear modes in higher order anisotropic flow ($V_{4}$, $V_{5}$ and $V_{6}$) in Pb--Pb collisions at $\sqrt{s_{\rm NN}} =$ 2.76 TeV using the ALICE detector at the Large Hadron Collider (LHC). A significant contribution from a non-linear mode is observed. A new observable, the non-linear response coefficient, is measured as well. The comparison to theoretical calculations provides crucial information on dynamic of the created system especially at the freeze-out conditions, which are poorly known from previous flow measurements. 

\end{abstract}

\end{frontmatter}


\section{Introduction}
\label{sec:intro}

The primary goal of the ultra-relativistic heavy-ion collisions program is to create and study the quark-gluon plasma (QGP), a state of matter whose existence under extreme conditions is predicted by quantum chromodynamics. 
The anisotropic flow is an important tool to achieve this goal. Usually it is quantified with flow coefficients $v_{n}$ and corresponding flow symmetry planes $\Psi_{n}$ in the Fourier series decomposition of the particle azimuthal distribution in the transverse plane~\cite{Voloshin:1994mz}.
The anisotropic flow $V_{n}$, defined as $V_{n} = v_{n} e^{i n \Psi_{n}}$, have been measured at the CERN Large Hadron Collider (LHC)~\cite{ALICE:2011ab, Adam:2016izf}. These measurements of $_{n}$ coefficients, combine with the hydrodynamic model calculations, provide compelling evidence that the created QGP matter appears to behave like an almost perfect fluid. 
Recently, it has been shown that the correlations between $v_n$ of different order carries more information about initial conditions and the properties of the QGP~\cite{ALICE:2016kpq, Zhou:2016eiz}. In particular, the correlations between anisotropic flow coefficients, investigated using symmetric cumulants, provide stricter constraints on initial conditions and shear viscosity over entropy density ratio, $\eta/s$, than measurements of individual $v_{n}$ alone~\cite{ALICE:2016kpq,Zhou:2016eiz,Bilandzic:2013kga}. Current model calculations are unable to describe quantitatively the measured $v_n$ correlations~\cite{ALICE:2016kpq,Zhu:2016puf}. 

It is known that $V_{2}$ and $V_{3}$ have linear contribution associated with the same order anisotropy coefficient $\varepsilon_{n}$. Higher order anisotropic flow $V_{n}$ with $n>$ 3 has contributions not only from the linear response of the system to $\varepsilon_{n}$, but also contributions from $\varepsilon_{2}$ and/or $\varepsilon_{3}$. 
For a single event, the higher order anisotropic flow $V_{n}$ with $n=$ 4, 5 and 6 can be decomposed into the so-called linear and the non-linear modes, according to
\begin{eqnarray}
V_{4} &=&  V_{4}^{\rm NL} + V_{4}^{\rm L} = \chi_{422} (V_{2})^2 +  V_{4}^{\rm L} , \label{eq:V4}\\
V_{5} &=&  V_{5}^{\rm NL} + V_{5}^{\rm L} = \chi_{532} V_{2} \, V_{3} + V_{5}^{\rm L} , \label{eq:V5}\\
V_{6} &=&  V_{6}^{\rm NL} + V_{6}^{\rm L} =  \chi_{6222} (V_{2})^3 + \chi_{633}(V_{3})^2 + \chi_{642} V_{2} V_{4}^{\rm L} + V_{6}^{\rm L}. \label{eq:V6}
\end{eqnarray}
Here $V_{n}^{\rm NL}$ and $V_{n}^{\rm L}$ represent the non-linear and linear contributions to the higher order anisotropic flow, whose magnitudes are denoted as $v_{n, ml}$ (or $v_{n, mlk}$) and $v_{n}^{\rm L}$, respectively. The ratio of $v_{n,ml}$ and $v_{n}\{2\}$, denoted as $\rho_{mn}$, could be used to probe the correlations between flow symmetry planes~\cite{Yan:2015jma}. Notice that $n$ stands for the order of anisotropic flow, $m$, $l$ and $k$ are the lower order of anisotropic flow. The $\chi_{mn}$ (or $\chi_{mnl}$) is a newly proposed observable called the non-linear mode coefficient which quantifies the contributions of non-linear model without contribution from $v_{2}$ and/or $v_3$~\cite{Yan:2015jma}. In these proceedings, we investigate both the linear and non-linear contributions to higher order anisotropic flow, with above mentioned observables.

\section{Analysis Details}

The data used in this analysis was recorded during the 2010 LHC Pb--Pb runs at $\sqrt{s_{_{\rm NN}}}=$ 2.76 TeV with the ALICE detector. For more details about the ALICE detector, we refer to~\cite{Aamodt:2008zz}. About 16 million events was recorded with a minimum-bias trigger, based on signals from two VZERO detectors ($-3.7\textless \eta \textless-1.7$ for VZERO-C and 2.8$\textless \eta \textless$5.1 for VZERO-A) and from the Silicon Pixel Detector, the innermost part of the Inner Tracking System (ITS, $|\eta|<0.8$). Charged tracks are reconstructed using the ITS and the Time Projection Chamber with full azimuthal coverage in the pseudo-rapidity range $|\eta|\textless$0.8. The 2- and multi-particle correlations are measured using the generic framework introduced in~\cite{Bilandzic:2013kga}.

\section{Results and Discussions}

\begin{figure}[tbh]
\begin{center}
\includegraphics*[width=14cm]{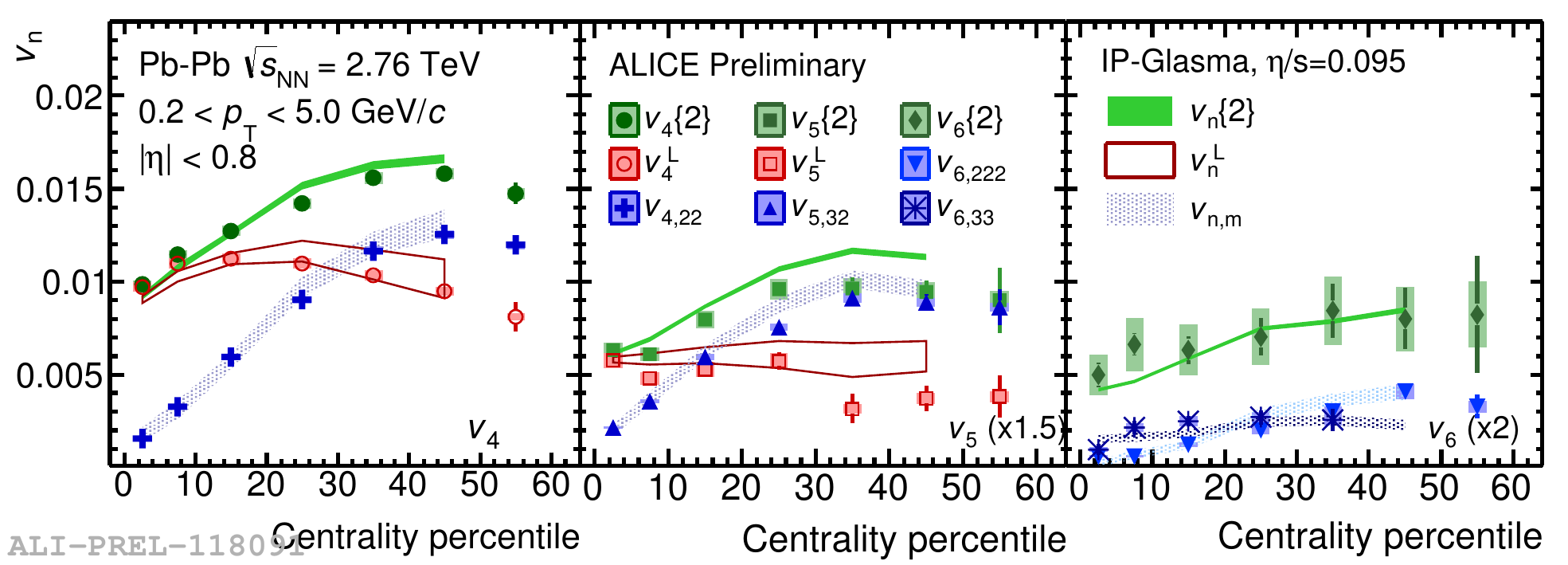}
\caption{
Centrality dependence of $v_4$ (left), $v_5$ (middle) and $v_6$ (right) in Pb--Pb collisions at $\sqrt{s_{_{\rm NN}}} = 2.76$ TeV. Contributions from linear and non-linear modes are presented with open and solid markers, respectively. The hydrodynamic calculations from ${\tt IP}$-${\tt Glasma+MUSIC+UrQMD}$~\cite{McDonald:2016vlt} are shown for comparison.
}
\label{fig:f1}
\end{center}
\end{figure}

The magnitudes of linear and non-linear modes in higher order anisotropic flow, $v_{n}^{\rm \,L}$ and $v_{n,ml}$, are shown in Fig.~\ref{fig:f1}. A pseudorapidity gap $|\Delta\eta|>0.8$ is applied for all measurements presented here. It is observed that $v_{4}^{\rm \, L}$ has a dominant contribution to $v_{4}\{2\}$ for the centrality range 0-30\% and it only changes modestly from central to peripheral collisions. Concerning $v_{4, 22}$, it increases significantly as a function of centrality and plays a dominant role for centrality classes above 30\%. Similar results can be observed for $V_{5}$. 
The physics for $V_6$ is a bit more complex, because it has several non-linear modes. However only $v_{6, 222}$ and $v_{6, 33}$ are discussed in these proceedings. As can be seen in Fig.~\ref{fig:f1} $v_{6, 222}$ has a strong dependence on centrality while $v_{6, 33}$ seems have a weakly dependence on centrality. 
All the above mentioned results are compared to the hybrid model ${\tt IP}$-${\tt Glasma+MUSIC+UrQMD}$~\cite{McDonald:2016vlt} calculations. Agreement is observed for $V_{4}$ and $V_{6}$, while the model calculations slightly overestimate the results for $V_{5}$.

\begin{figure}[tbh]
\begin{center}
\includegraphics*[width=11cm]{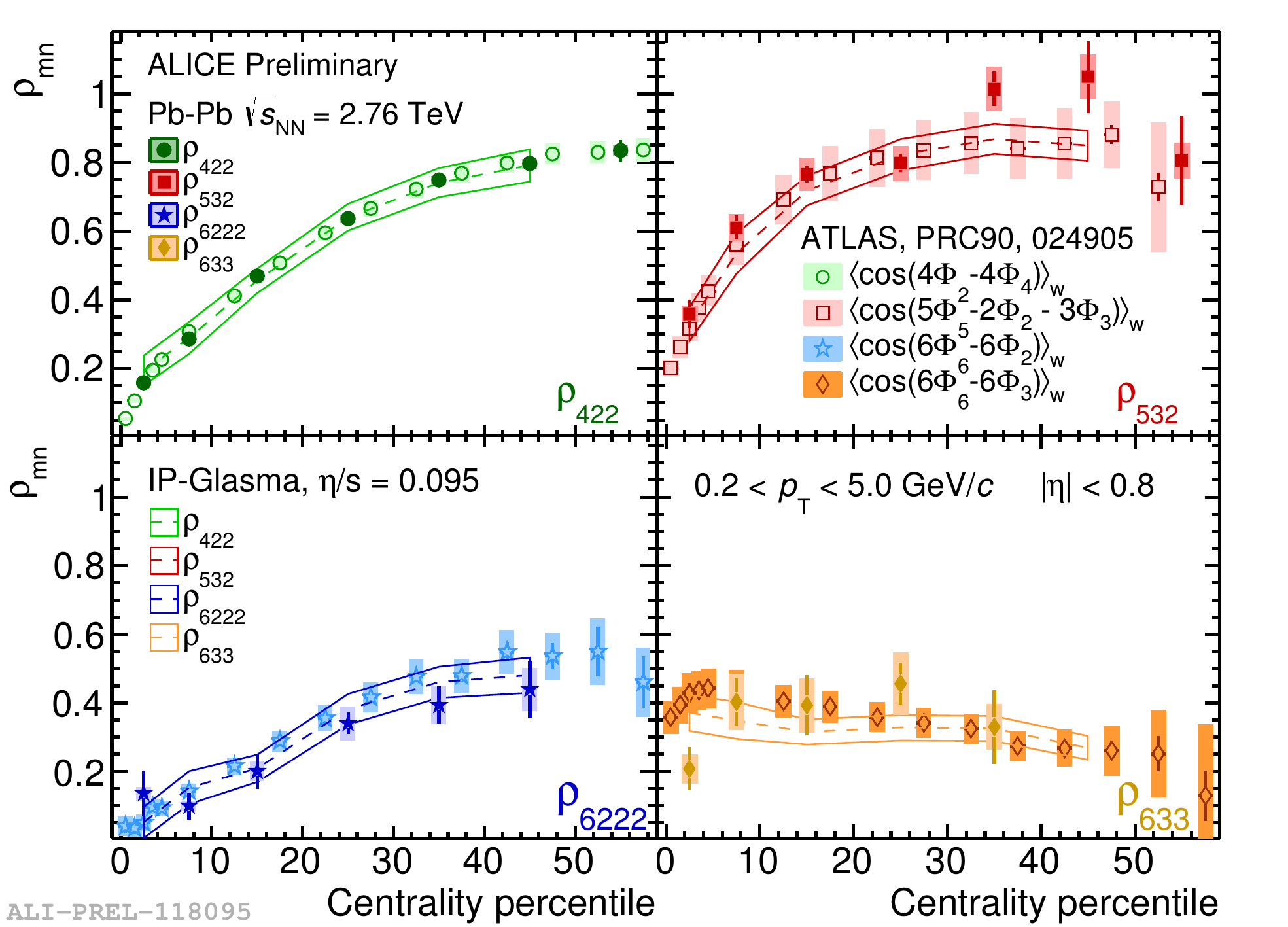}
\caption{
Centrality dependence of $\rho_{mn}$ in Pb--Pb collisions at $\sqrt{s_{_{\rm NN}}} = 2.76$ TeV.  ATLAS measurements based on the event-plane correlation~\cite{Aad:2014fla} are presented with open markers. The hydrodynamic calculations from ${\tt IP}$-${\tt Glasma+MUSIC+UrQMD}$~\cite{McDonald:2016vlt} are shown with open bands.
}
\label{fig:f2}
\end{center}
\end{figure}

Figure~\ref{fig:f2} shows the centrality dependence of $\rho_{mn}$, which is defined as the ratio of $v_{n,ml}$ and $v_{n}\{2\}$. This observable measures the flow symmetry plane correlations. It can be seen that $\rho_{422}$, $\rho_{532}$ and $\rho_{6222}$ increase as centrality increases. The results of $\rho_{633}$ do not exhibit a strong centrality dependence, considering the larger statistical uncertainties. 
This measurement is also compared to the ``event-plane correlation'' measurements from ATLAS Collaboration~\cite{Aad:2014fla}. 
Although the kinematic cuts used by ATLAS and this analysis are different, the results are compatible with each other. 
It is also shown in Fig.~\ref{fig:f2} that the calculations from ${\tt IP}$-${\tt Glasma+MUSIC+UrQMD}$~\cite{McDonald:2016vlt} could describe quantitatively the $\rho_{mn}$ results. 
It was suggested by both hydrodynamic~\cite{Qiu:2012uy} and transport model~\cite{Zhou:2015eya} calculations that stronger initial participant plane correlations or a smaller value of $\eta/s$ of the QGP lead to a stronger correlations between the flow symmetry planes. Therefore, the measurements of $\rho_{mn}$ presented in these proceedings provide constraints on both the initial conditions and $\eta/s$ of the QGP in model calculations. 

\begin{figure}[tbh]
\begin{center}
\includegraphics*[width=12cm]{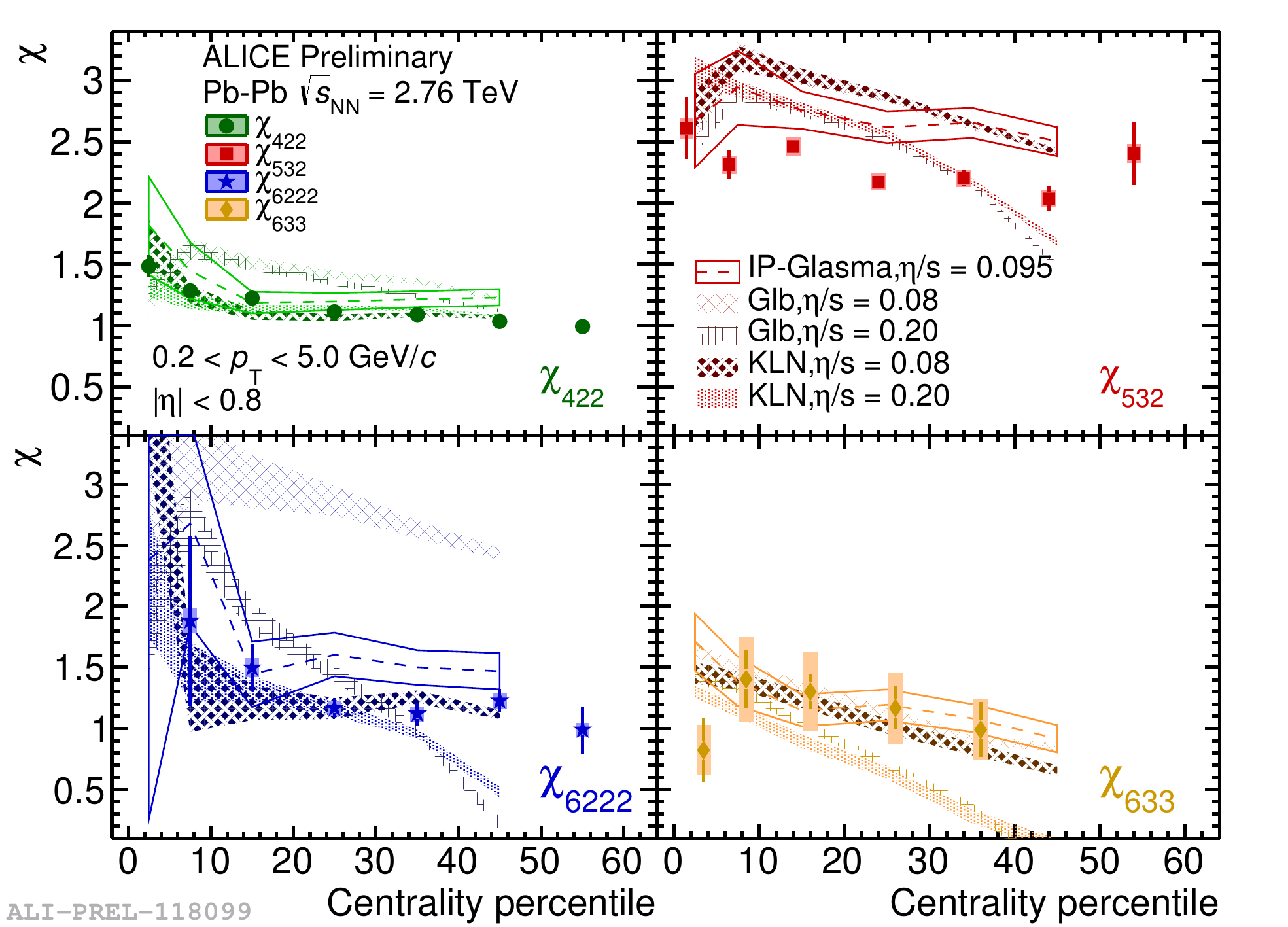}
\caption{
Centrality dependence of $\chi$ in Pb--Pb collisions at $\sqrt{s_{_{\rm NN}}} = 2.76$ TeV. Hydrodynamic calculations from ${\tt VISH2+1}$~\cite{Qian:2016fpi} are shown in shaded areas and the one from ${\tt IP}$-${\tt Glasma+MUSIC+UrQMD}$~\cite{McDonald:2016vlt} are shown with open bands.
}
\label{fig:f3}
\end{center}
\end{figure}

The measurements of non-linear mode coefficients $\chi_{mn}$ are presented in Fig.~\ref{fig:f3}. The $\chi_{422}$ and $\chi_{6222}$ show a weak decrease from central to mid-central collisions, and stay constant toward more peripheral collisions. $\chi_{633}$ seems to have a decreasing trend towards peripheral collisions within large uncertainties, whereas $\chi_{532}$ is almost independent on centrality. Therefore, the increase of $v_{n,ml}$ as a function of centrality percentile presented in Fig.~\ref{fig:f1} could be mainly explained by the increase of $v_{2}$ and/or $v_{3}$ as centrality increases, not by the increase of the non-linear mode coefficient $\chi_{mn}$. We notice that the hydrodynamic prediction~\cite{Yan:2015jma} of $\chi_{422} \sim \chi_{633} \sim \frac{\chi_{532}}{2}$ are reproduced by the data.
Figure~\ref{fig:f3} also shows the calculations of viscous hydrodynamics from ${\tt VISH2+1}$~\cite{Qian:2016fpi} and hybrid model ${\tt IP}$-${\tt Glasma+MUSIC+UrQMD}$~\cite{McDonald:2016vlt}. 
The data-theory comparison presented here indicate that the data favors the descriptions using IP-Glasma and MC-KLN over MC-Glauber initial conditions regardless of $\eta/s$.

\section{Summary}

In summary, we present the studies on the linear and non-linear contributions to the higher order anisotropic flow in Pb--Pb collisions at $\sqrt{s_{_{\rm NN}}}=$ 2.76 TeV. Measurements and the comparisons to hydrodynamic calculations offer new insights into the geometry of the fluctuating initial state and provide further understanding of the dynamical evolution of the strongly interacting medium produced in relativistic heavy-ion collisions at the LHC.


\bibliographystyle{elsarticle-num}
\bibliography{<your-bib-database>}



\end{document}